\documentclass[runningheads]{llncs}
\usepackage[T1]{fontenc}
\usepackage{graphicx}
\usepackage{pifont}

\usepackage{graphicx}

\usepackage{amsmath,mathtools,amssymb,amsfonts,bm}

\usepackage{subcaption}
\usepackage{graphicx}
\usepackage{caption}
\usepackage{booktabs}
\usepackage{multirow}
\usepackage{hyperref}
\hypersetup{
    colorlinks=true,
    linkcolor=blue,
    filecolor=cyan,      
    urlcolor=magenta,
    }
\usepackage[nameinlink]{cleveref}
\usepackage{color}
\usepackage[table]{xcolor}
\usepackage{arydshln}
\usepackage{makecell}
\usepackage{svg}

\usepackage{wrapfig}
\usepackage{url}
\usepackage{array}

\definecolor{myorange}{rgb}{1, 0.647, 0}
\definecolor{myblue}{rgb}{.118, 0.565, 1}
\begin{document}
\title{Segmentation of Brain Metastases in MRI: A Two-Stage Deep Learning Approach with Modality Impact Study
}
%
%
\author{
Yousef Sadegheih\inst{1}\orcidID{0009-0003-1766-5519} \and
Dorit Merhof\inst{1,2}\orcidID{0000-0002-1672-2185}
}
%
\authorrunning{Y. Sadegheih and D. Merhof}
%
\institute{
Faculty of Informatics and Data Science, University of Regensburg, Regensburg, 93053, Germany \and
Fraunhofer Institute for Digital Medicine MEVIS, Bremen 28359, Germany\\ 
\email{ \{dorit.merhof@ur.de\}}
}
\maketitle              
\begin{abstract}
Brain metastasis segmentation poses a significant challenge in medical imaging due to the complex presentation and variability in size and location of metastases. In this study, we first investigate the impact of different imaging modalities on segmentation performance using a 3D U-Net. Through a comprehensive analysis, we determine that combining all available modalities does not necessarily enhance performance. Instead, the combination of T1-weighted with contrast enhancement (T1c), T1-weighted (T1), and FLAIR modalities yields superior results. Building on these findings, we propose a two-stage detection and segmentation model specifically designed to accurately segment brain metastases. Our approach demonstrates that leveraging three key modalities (T1c, T1, and FLAIR) achieves significantly higher accuracy compared to single-pass deep learning models. This targeted combination allows for precise segmentation, capturing even small metastases that other models often miss. Our model sets a new benchmark in brain metastasis segmentation, highlighting the importance of strategic modality selection and multi-stage processing in medical imaging.  Our implementation is freely accessible to the research community on \href{https://github.com/xmindflow/Met-Seg}{GitHub}.


\end{abstract}

\section{Introduction}
\label{sec:introduction}

Brain metastases are increasingly relevant in oncology, occurring when cancer cells spread to the brain from other parts of the body. This condition affects 20-40\% of cancer patients and contributes to high morbidity and mortality. Accurate detection and treatment are therefore critical to improving patient outcomes~\cite{pfluger2022automated,mehrabian2019advanced,jeong2024deep}. While advances in medical imaging, particularly magnetic resonance imaging (MRI), have enhanced our ability to acquire images of brain metastases at different contrasts, accurately segmenting these metastases remains a challenging task~\cite{yin2022development}. 
Traditional methods for segmenting brain metastases rely on manual delineation by radiologists. However, this process is time-consuming and prone to variability~\cite{xue2020deep,perez2016brain}. Recently, deep learning techniques have shown significant promise in automating brain metastasis segmentation. These techniques have transformed medical image analysis, providing effective tools for various tasks, including segmentation~\cite{yao2024cnn}. Most current methods use a single model to segment brain metastases directly from MRI scans~\cite{grovik2020deep,ozkara2023deep,yang20243d}. For instance, MLP-UNEXT combines Convolutional Neural Networks (CNNs) with Multi-Layer Perceptrons (MLPs) to segment brain metastases~\cite{li2024mlp}, a modified V-Net 3D CNN focuses on patients treated with Stereotactic Radiosurgery (SRS) using both MRI and CT scans~\cite{hsu2021automatic}, and a modified GoogLeNet architecture has been employed for single-pass segmentation of brain metastases~\cite{grovik2020deep}. Additionally, 3D-TransUNet~\cite{yang20243d} employs a hybrid model of transformers and CNNs using pre-trained Masked-Autoencoder (MAE) and deep supervision. This model utilizes the Hungarian matching loss to establish correspondence between predictions and ground-truth segments but requires extensive pre-training.
Moreover, Erdur et al.~\cite{erdur2023all} modified the SegResNetVAE architecture and introduced a blob loss to improve metastasis segmentation. These studies, including 3D-TransUNet and Erdur et al., leverage the BraTS-Mets 2023 dataset~\cite{moawad2023brain} with extensive pre-training and auxiliary datasets to enhance performance.
While these methods have demonstrated success, they also face challenges, such as detecting very small metastases or distinguishing them from surrounding tissue. Furthermore, there is inconsistency in the imaging modalities used by these models, which can affect their performance.\\
\indent In this paper, we first investigate the impact of different MRI modalities on segmentation performance. MRI offers various imaging sequences, such as T1-weighted (T1), T1-weighted with contrast-enhanced (T1c), T2-weighted (T2), and Fluid Attenuated Inversion Recovery (FLAIR), each providing unique information about the brain's structure and pathology. Understanding how each modality influences segmentation accuracy can guide the development of more effective and specialized deep-learning models for this task.
Building on these insights, we propose a two-stage approach to improve segmentation accuracy for brain metastases. Our method consists of two sequential deep-learning models. The first model is a patch-based detector trained to identify small regions within the MRI scans that potentially contain metastases. This step significantly reduces the search space for the subsequent segmentation model, focusing its efforts on areas most likely to contain metastases. The second model then performs detailed segmentation within these identified patches, allowing for more precise delineation of metastases.
In summary, the key contributions of this paper are as follows:
\ding{182} We conduct a comprehensive evaluation of various MRI modalities to assess their impact on segmentation performance.
\ding{183} We propose a two-stage segmentation approach that separates the detection of metastasis-containing patches from the segmentation of metastases within these patches, aiming to improve segmentation accuracy and efficiency using just three modalities instead of four modalities.
\ding{184} We provide a detailed comparison with state-of-the-art single-pass segmentation models, demonstrating the advantages of our method in handling small and heterogeneous brain metastases.
Through this work, we aim to advance the field of automated brain metastasis segmentation and provide valuable insights into the benefits of a two-stage approach and the role of different MRI modalities in this context.

\begin{figure}[!t]
    \centering
    \resizebox{\textwidth}{!}{
        \includegraphics[width=0.9\textwidth]{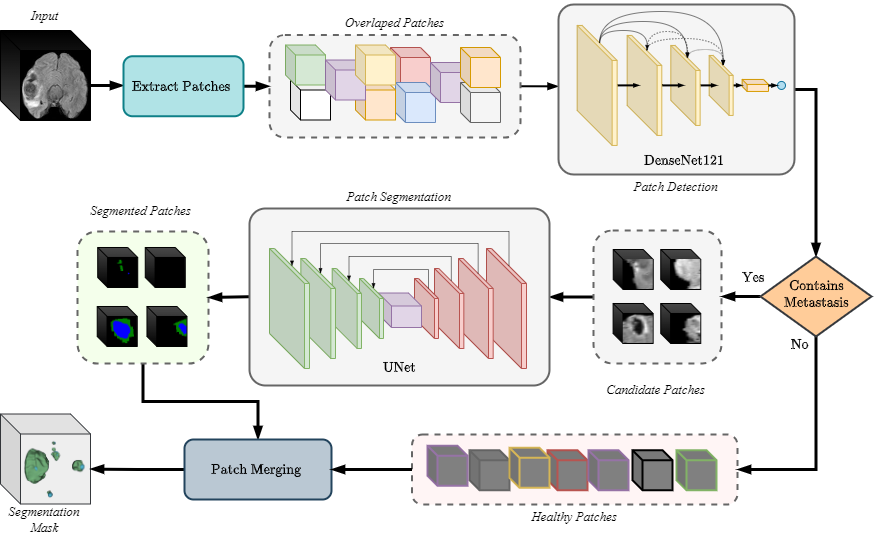}
}
    \caption{Workflow of two-stage model architecture for detection and segmentation of the brain metastasis.} 
    
    \label{fig:model}
\end{figure}

\section{Methodology}
\label{sec:method}
\Cref{fig:model}  shows our two-stage approach for segmenting brain metastases. The first stage detects regions with potential metastases, and the second stage segments these regions. Below, we explain each stage in more detail.

\subsection{Detector}
To make the segmentation process more efficient, we use DenseNet121~\cite{huang2017densely} to identify regions that may contain metastases. DenseNet121 is chosen due to its good balance between accuracy and computational efficiency, considering the number of trainable parameters and floating point operations per second (FLOPS). Its dense connections help with gradient flow and prevent gradient vanishing, which is important in training deep networks. The detector works by analyzing overlapping patches from the input images. We keep the patch sizes small to ensure that smaller metastases are not missed due to larger ones influencing the region of interest.

The detection stage offers significant benefits. It reduces the area that the segmentation model needs to analyze, allowing more computational power to focus on relevant regions. This improves both accuracy and efficiency, leading to better identification of metastasis boundaries. By targeting specific areas, we can use computational resources more effectively, which is especially important in large-scale medical imaging where processing power and time are crucial. This approach speeds up the segmentation process. The method ensures that small metastases are detected early, improving sensitivity and overall management of brain metastases. The initial detection stage filters the data, allowing the segmentation model to concentrate on important regions, enhancing accuracy and precision in treatment planning. 

\subsection{Segmentor}
For segmenting the patches identified by the detector, we use a modified 3D-UNet~\cite{cciccek20163d,isensee2021nnu}. Our version includes residual connections in each layer and deep supervision. Residual connections help with gradient flow, stabilizing the training. Deep supervision gives feedback to intermediate layers, helping the network learn better features. These changes let the model focus on the important parts needed for accurate segmentation. By concentrating on the regions identified by the detector, the segmentation model can better identify and segment the metastases.

The segmentation stage has several benefits. Residual connections enhance information flow within the network, aiding in detailed feature learning. This is crucial for detecting the varied characteristics of brain metastases. Deep supervision helps learn features at multiple levels, aiding in the accurate segmentation of metastases of different sizes. Focusing on specific regions identified by the detector reduces false positives and improves accuracy. This leads to better identification of metastases, which is important for treatment planning. Accurate segmentation helps clinicians determine the size and location of metastases, which is crucial for deciding the best treatment.
Additionally, targeting specific areas for segmentation makes the model more efficient, reducing computational load and speeding up processing time. We also use region-based optimization to ensure consistent and accurate segmentation across different regions, enhancing the reliability of our results.

In summary, the modifications to the 3D-UNet, including residual connections and deep supervision, along with region-based optimization, enable more effective segmentation of brain metastases. These improvements help capture a wide range of metastasis characteristics, reduce false positives, and enhance overall segmentation performance. Our two-stage approach uses a DenseNet121-based detector to find potential metastasis regions, followed by a modified 3D-UNet for detailed segmentation. This setup improves computational efficiency and accuracy, making the results more reliable.

\section{Experimental setup and analysis}
\label{sec:experiment}
\subsection{Dataset}
\label{subsec:dataset}
\begin{figure}[!t]
    \centering
    \resizebox{\textwidth}{!}{
        \begin{tabular}{@{} *{3}c @{}}
            \includegraphics[width=1\textwidth]{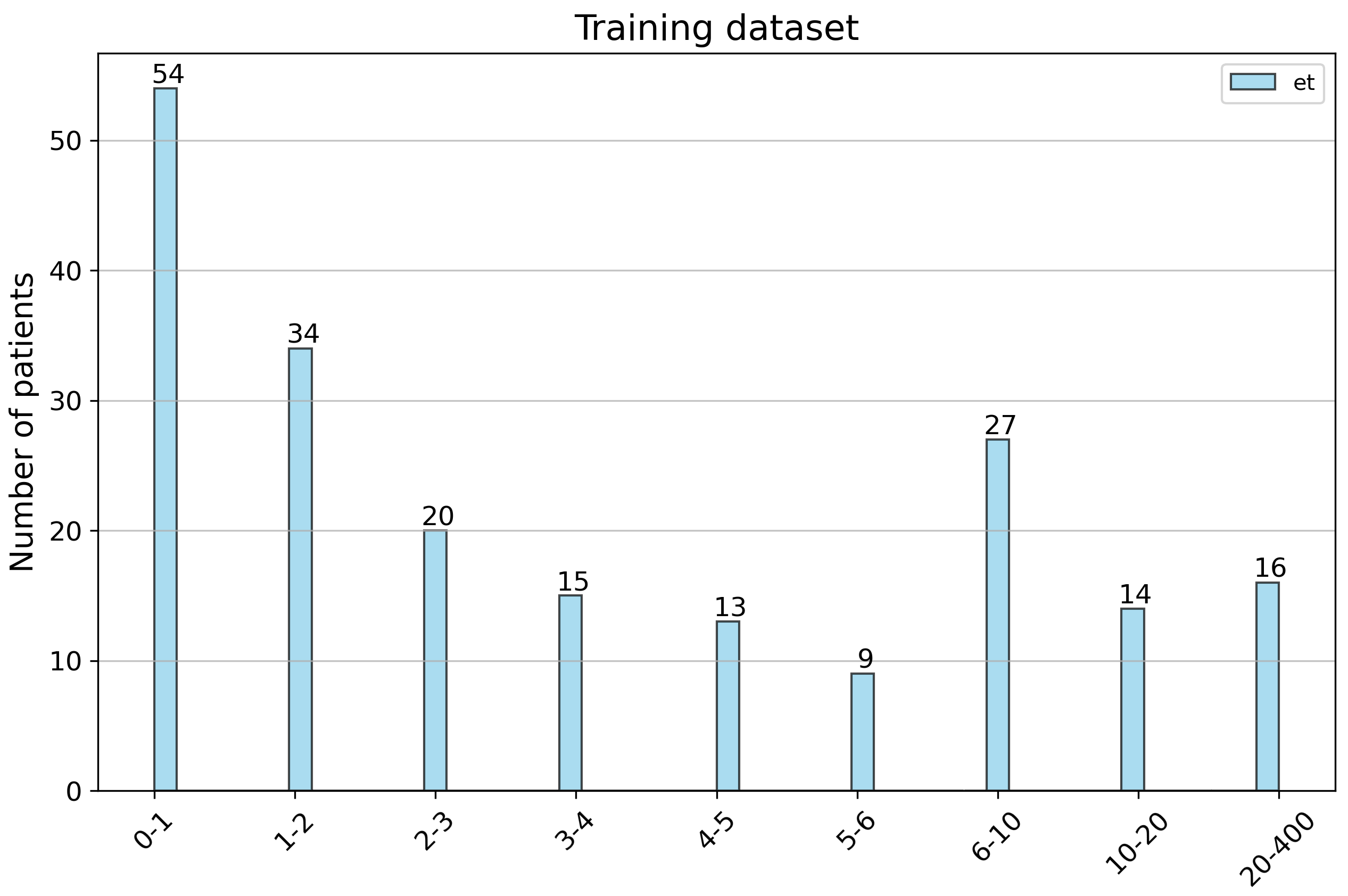} &
            \includegraphics[width=1\textwidth]{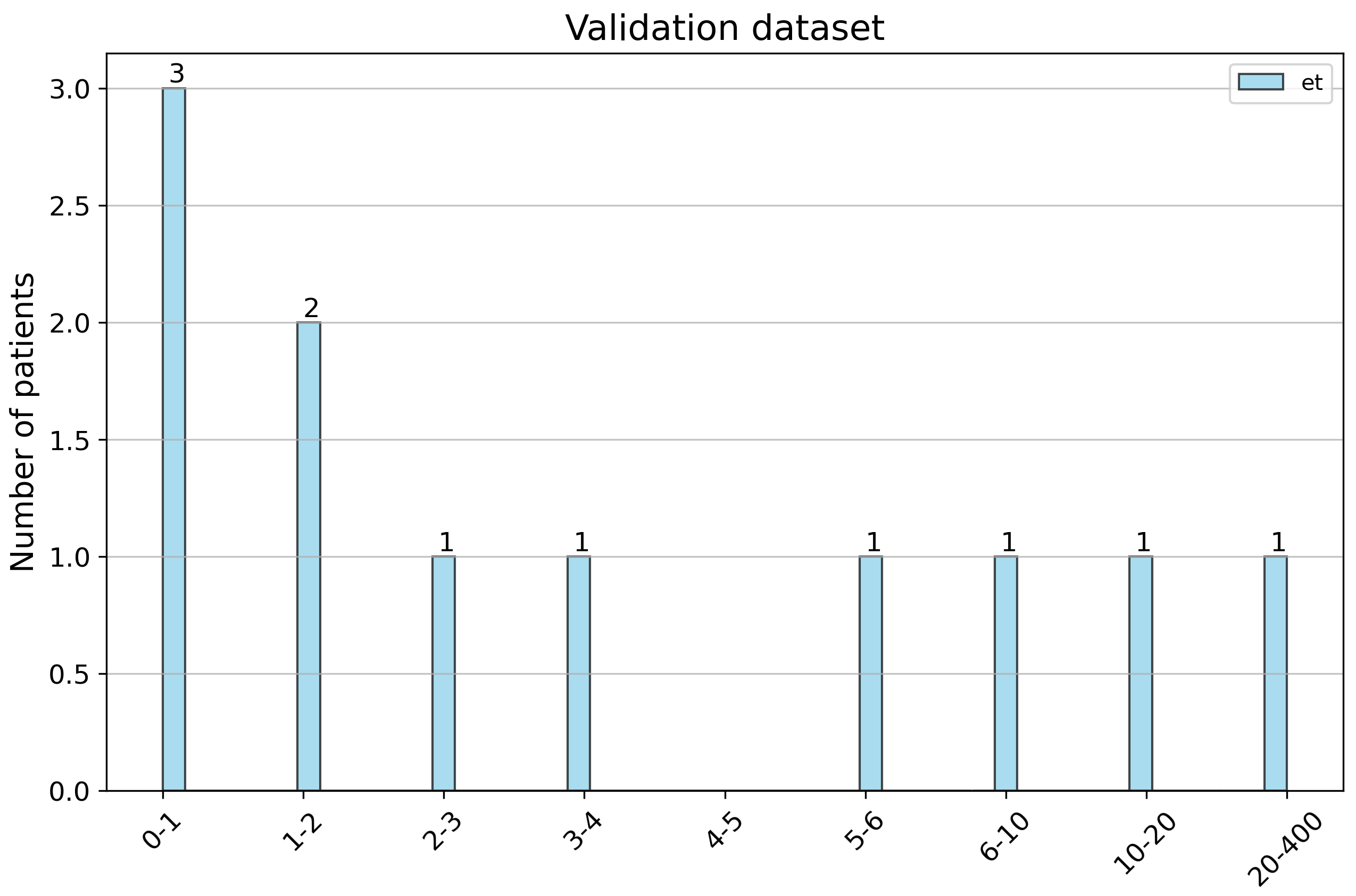} &
            \includegraphics[width=1\textwidth]{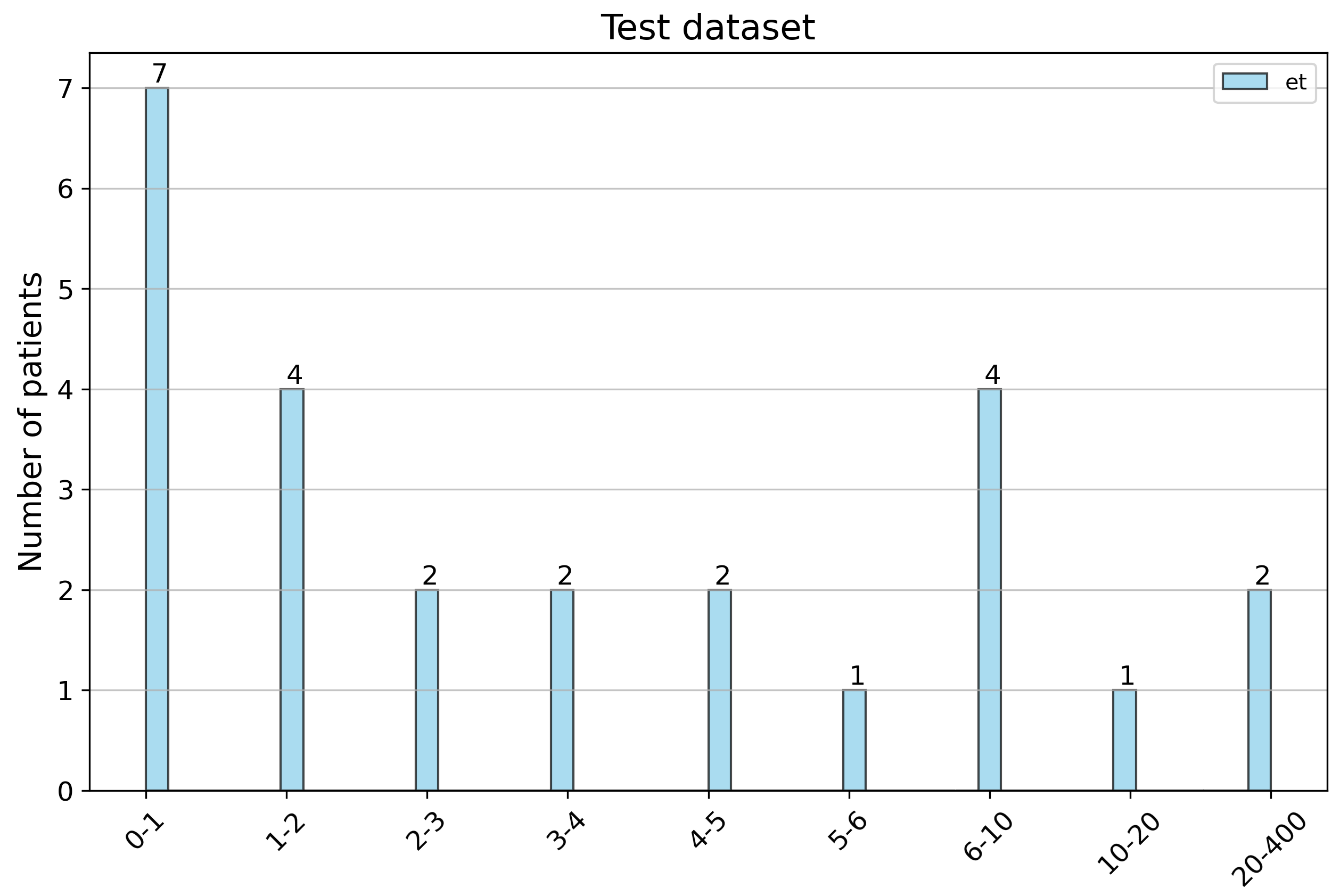}\\
        \end{tabular}    }
    \caption{Distribution of patients in the training, validation, and test sets based on the number of metastases in the BraTS-Mets 2023 dataset. The figure illustrates how many patients fall into different categories of metastasis count.} 
    
    \label{fig:data_distribution}
\end{figure}

To evaluate our method, we used the BraTS-Mets 2023 dataset~\cite{moawad2023brain}. This dataset has MRI scans from 238 patients with four MRI modalities: T1, T1c, T2, and FLAIR. Annotations include non-enhancing tumor core (NETC; Label 1), surrounding non-enhancing FLAIR hyperintensity (SNFH; Label 2), and enhancing tumor (ET; Label 3). Tumor core (TC) is ET + NETC, and whole tumor (WT) is all labels. Scans are resampled to 1x1x1 mm, 240x240x155 size. The dataset is split into 202 patients ($\sim$85\%) for training, 11 ($\sim$5\%) for validation, and 25 ($\sim$10\%) for testing. Data distribution is shown in~\Cref{fig:data_distribution}.

\subsection{Experimental setup}
\label{subsec:setup}
The framework was developed using Python 3.10.12 and PyTorch 2.1.0~\cite{paszke2019pytorch,Falcon_PyTorch_Lightning_2019}. The model was trained on a single NVIDIA A100 GPU with 80 GB of VRAM. Binary cross-entropy was used for detection loss, while a combination of DICE loss~\cite{milletari2016v} and binary cross-entropy with deep supervision was used for segmentation loss: \(\mathcal{L}_{total} = \mathcal{L}_{dice} + \mathcal{L}_{bce}\).

For segmentation, SGD with Nesterov momentum was used. The settings included a weight decay of $3 \times 10^{-5}$, momentum of 0.99, and an initial learning rate of 0.01, updated each epoch: \(\text{learning rate} = \text{base } lr \times \left( \frac{\text{epoch}}{\text{max\_epoch}} \right)^{0.9}\). Adam~\cite{Kingma2014AdamAM} was used for detection with a learning rate of 0.0004, \(\beta_{1} = 0.9\), and \(\beta_{2} = 0.999\). A linear warm-up cosine annealing scheduler was used, starting with a learning rate of $1 \times 10^{-6}$ and warming up for 10 epochs.

\sloppy
Data augmentation included random rotation, scaling, Gaussian noise, smoothing, mirroring, and random simulation of low resolution~\cite{isensee2021nnu}. The detection model was trained for 400 epochs with a batch size of 2 and 5 random crop samples per patient. The patch size was 64x64x64. The segmentation model was trained for 250,000 iterations with the same batch size and patch size. To ensure robustness, all runs were performed three times with different random seeds.

\subsection{Metrics}
\label{subsec:metrics}
We use both the traditional Dice Similarity Coefficient (DSC) and 95\% Hausdorff Distance (HD95), along with Lesion-wise DSC and HD95~\cite{moawad2023brain,Chung2023}. Lesion-wise metrics are calculated as:

\begin{equation}
\text{Lesion-wise Metric} = \frac{\sum_{i}^{L} \text{Metric}(l_i)}{TP + FN + FP}
\end{equation}

\noindent where \(L\) is the number of ground truth lesions, \(l_i\) is the pair of ground truth and its corresponding prediction, and "Metric" is either HD95 or DSC. This method penalizes false negatives and positives, which is important in clinical practice. Lesions smaller than 2 voxels are ignored.

\begin{table*}[!t]
\renewcommand{\r}[1]{\textcolor{red}{#1}}
\renewcommand{\b}[1]{\textcolor{blue}{#1}}
    \centering
    \caption{Impact of different imaging types on DSC in brain metastasis segmentation using a 3D U-Net. The table shows the performance of T1, T1c, T2, and FLAIR, both alone and in combinations. WT is Whole Tumor, TC is Tumor Core, NETC is Non-Enhancing Tumor Core, and SNFH is Surrounding Non-Fluid Hyperintensities. Mean DSC and Lesion-wise DSC metrics are reported with standard deviations in parentheses. Each model was trained three times with different seeds. \b{\textbf{Blue}} indicates the best result, and \r{\textbf{red}} displays the second-best.}
    \resizebox{\textwidth}{!}{
    \begin{tabular}{l||cccc|cc||cccc}
\hline
\multirow{2}{*}{Modalities} & \multicolumn{6}{c||}{Legacy DSC \(\uparrow\)} & \multicolumn{4}{c}{Lesion-wise DSC \(\uparrow\)} \\ \cline{2-11} 
                            & WT & TC & ET & {AVG.} & NETC & SNFH & WT     & TC     & ET     & AVG.     \\ \hline
                            \rowcolor{gray!5}
      t1          & 42.43 (2.02)	    &   32.05 (3.06)        &   26.63 (3.12)        &   \multicolumn{1}{c|}{33.70 (2.72)}	        &       5.62 (1.55)	        &   33.86 (1.19)	    &   24.13 (2.16)	    &   15.62 (1.12)	    &   12.46 (1.00)	    &   17.40 (1.42) \\
                           \rowcolor{gray!10}
     t1c         & 64.74 (3.11)      &   74.81 (2.89)	    &   69.02 (2.81)	    &   \multicolumn{1}{c|}{69.52 (2.94)}	        &       \b{42.01 (2.64)}	&   44.07 (1.75)	    &   43.69 (4.55)	    &   \b{53.19 (1.82)}	&   \b{48.98 (1.97)}	&   \r{48.62 (2.74)} \\
                            \rowcolor{gray!5}
      t2          & 59.50 (1.82)      &   46.39 (2.01)        &	38.28 (0.61)        &	\multicolumn{1}{c|}{48.06 (1.44)}           &	    6.35 (0.85)         &	45.83 (0.69)        &	35.26 (1.68)        &	26.80 (0.53)        &	21.57 (0.30)        &	27.88 (0.64) \\
                            \rowcolor{gray!10}
      f           & 69.76 (1.72)      &	47.97 (2.74)        &	39.49 (2.31)        &	\multicolumn{1}{c|}{52.41 (2.25)}           &	    4.04 (2.28)         &	53.59 (1.75)        &	36.62 (4.62)        &	27.33 (0.76)        &	22.33 (0.71)        &	28.76 (1.18) \\
                            \midrule
                            \rowcolor{gray!5}
       t1+t1c      & 65.69 (2.88)      &	\r{75.35 (2.81)}    &	\r{69.75 (3.05)}    &	\multicolumn{1}{c|}{70.26 (2.91)}           &	    \r{39.80 (3.43)}    &	43.41 (1.11)        &	40.71 (2.73)        &	51.95 (4.59)        &	47.53 (4.93)        &	46.73 (4.07) \\
                            \rowcolor{gray!10}
       t1+t2       & 58.46 (2.29)      &	44.09 (2.52)        &	35.47 (1.33)        &	\multicolumn{1}{c|}{46.01 (1.91)}           &	    8.27 (1.08)         &	45.67 (1.37)        &	33.47 (2.05)        &	26.61 (0.56)        &	21.08 (0.50)        &	27.05 (0.77)\\
                            \rowcolor{gray!5}
      t1+f        & 62.82 (1.58)      &	42.46 (1.67)        &	35.62 (2.06)        & 	\multicolumn{1}{c|}{46.96 (1.77)}           &	    3.10 (1.38)         &	49.52 (0.98)        &	35.92 (4.17)        &	25.05 (0.21)        &	20.58 (0.27)        &	27.18 (1.34) \\
                          \rowcolor{gray!10}
      t1c+t2      & 70.53 (1.39)      &	74.07 (1.28)        &	68.34 (1.06)        &	\multicolumn{1}{c|}{70.98 (1.21)}           &	    38.35 (3.17)        &	51.92 (1.49)        &	44.30 (2.90)        &	48.63 (2.29)        &	44.38 (2.34)        &	45.77 (1.84) \\
                            \rowcolor{gray!5}
      t1c+f       & 69.72 (2.17)      &	70.37 (2.50)        &	65.24 (2.19)        &	\multicolumn{1}{c|}{68.45 (2.23)}           &	    38.55 (2.07)        &	53.08 (1.26)        &	\r{47.95 (4.41)}    &	48.19 (1.93)        &	44.46 (1.75)        &	46.87 (2.69) \\
                            \rowcolor{gray!10}
      t2+f        & 64.15 (2.11)      &	44.36 (3.85)        &	37.23 (2.52)        &	\multicolumn{1}{c|}{48.58 (2.82)}           &	    4.89 (1.11)         &	48.40 (0.61)        &	36.19 (3.00)        &	23.88 (4.14)        &	19.44 (3.22)        &	26.50 (3.15)\\
                            \midrule
                            \rowcolor{gray!5}
     t1+t1c+t2   & 68.66 (1.63)      &	\b{75.78 (0.65)}    &	\b{70.17 (1.02)}    &	\multicolumn{1}{c|}{71.54 (1.10)}           &	    39.56 (1.23)        &	48.87 (2.39)        &	46.99 (1.30)        &	48.40 (5.36)        &	44.79 (4.96)        &	46.73 (3.87)\\
                            \rowcolor{gray!10}
     t1+t1c+f   & \b{73.02 (0.83)}  &	74.07 (1.42)        &	69.55 (1.41)        &	\multicolumn{1}{c|}{\b{72.21 (1.17)}}       &	    35.28 (0.69)        &	54.33 (1.15)        &	\b{48.47 (0.94)}    &	51.12 (4.63)        &	\r{47.61 (4.02)}    &	\b{49.07 (3.16)}\\
                            \rowcolor{gray!5}
    t1+t2+f     & 63.21 (2.93)      &	44.67 (2.60)        &	37.61 (2.38)        &	\multicolumn{1}{c|}{48.50 (2.63)}           &	    4.84 (1.47)         &	49.34 (2.49)        &	33.85 (4.90)        &	24.71 (2.06)        &	20.93 (2.23)        &	26.50 (3.00)\\
                            \rowcolor{gray!10}
     t1c+t2+f   & \r{72.69 (0.87)}  &	73.76 (1.39)        &	68.50 (0.86)        &	\multicolumn{1}{c|}{\r{71.65 (0.44)}}       &	    34.70 (0.35)        &	\b{55.37 (2.03)}    &	45.13 (3.95)        &	\r{51.99 (1.62)}    &	47.32 (0.95)        &	48.15 (0.93) \\

    \midrule
    \rowcolor{gray!5}
    \textbf{t1+t1c+t2+f}                & 71.61 (0.46)     &	71.41 (0.37)        &	66.50 (0.98)        &	\multicolumn{1}{c|}{69.84 (0.60)}           &	    27.73 (2.14)        &	\r{54.92 (0.87)}    &	44.41 (0.15)        &	49.15 (1.82)        &	44.99 (0.84)        &	46.18 (0.83) \\ 
    \bottomrule
    \end{tabular}
    }
\label{tab:results-modalities}
\end{table*}

\section{Results}
\label{sec:results}
\subsection{Quantitative results}
\label{subsec:quantitative}

\begin{table*}[!ht]
\renewcommand{\r}[1]{\textcolor{red}{#1}}
\renewcommand{\b}[1]{\textcolor{blue}{#1}}
    \centering
    \caption{Comparison of our method with state-of-the-art models on the BraTS Mets 2023 dataset. Top results are in \b{\textbf{blue}}, second best in \r{\textbf{red}}. Metrics include DSC and HD95 for the whole tumor (WT), tumor core (TC), and enhanced tumor (ET). Models were trained three times with different seeds. Mean values are reported with standard deviations below them. All models use a 128x128x128 patch size, except ours, which uses 64x64x64.}

    \resizebox{\textwidth}{!}{
    \begin{tabular}{l||cccccccc||cccccccc}

            \hline
    \multirow{3}{*}{Models} & \multicolumn{8}{c||}{Legacy Metrics}                                  & \multicolumn{8}{c}{Lesion-wise Metrics}                             \\ \cline{2-17} 
                        & \multicolumn{4}{c|}{DSC \(\uparrow\)}                 & \multicolumn{4}{c||}{HD95 \(\downarrow\)} & \multicolumn{4}{c|}{DSC \(\uparrow\)}                 & \multicolumn{4}{c}{HD95 \(\downarrow\)} \\ \cline{2-17} 
                        & WT & TC & ET & \multicolumn{1}{c|}{AVG.} & WT   & TC   & ET  & AVG.  & WT & TC & ET & \multicolumn{1}{c|}{AVG.} & WT   & TC  & ET  & AVG.  \\ \hline
                        ResUNet~\cite{lee2017superhuman}        &   72.52      & 73.01      & 67.01      & \multicolumn{1}{c|}{70.85}    & 23.93      & 28.22     & 27.72      & 26.62    & 46.10      & \r{49.53}      & 44.79      & \multicolumn{1}{c|}{46.81}    & 139.58     & 129.23       & 140.02     & 136.28 \\
                        & (0.73)       & (1.15)       & (0.95)       & \multicolumn{1}{c|}{(0.89)}     & (5.43)      & (6.72)       & (6.71)       & (6.28)     & (2.98)       & \r{(1.16)}       & (2.29)       & \multicolumn{1}{c|}{(0.19)}     & (19.65)      & (8.40)         & (6.56)       & (9.11) \\
                        \hdashline
                        ResUNetSE~\cite{toubal2020deep}         &   69.55      & 72.05      & 68.84      & \multicolumn{1}{c|}{70.15}    & 29.34         & 33.17         & 26.20   & 29.57      & 39.02      & 44.40      & 41.64      & \multicolumn{1}{c|}{41.69}    & 174.23     & 161.88        & 162.04     & 172.21 \\ 
                        &  (1.26)        &	 (0.50)     &	 (0.49) &	 \multicolumn{1}{c|}{(0.42)} &	 (7.17) &	 (4.85) &	 (4.81) &	 (5.61) &	 (0.93) &	 (2.51) &	 (1.96) &	 \multicolumn{1}{c|}{(1.80)} &	 (14.38) &	 (16.72) &	 (15.69) &	 (15.60) \\ 
                         \hdashline
                        3D-UNET~\cite{cciccek20163d}            &   71.61      & 71.41      & 66.50      & \multicolumn{1}{c|}{69.84}    & 25.64         & 28.85         & 29.06      & 27.85      & 44.41      & 49.15      & 44.99      & \multicolumn{1}{c|}{46.18}    & 151.75     & 137.58        & 146.29     & 145.21 \\ 
                          &  (0.46) &	 (0.37) &	 (0.98) &	 \multicolumn{1}{c|}{(0.60)} &	 (1.28) &	 (0.51) &	 (0.41) &	 (0.12) &	 (0.15) &	 (1.82) &	 (0.84) &	 \multicolumn{1}{c|}{(0.83)} &	 (5.70) &	 (14.05) &	 (13.77) &	 (11.18)\\ 
                         \hdashline
                        VNet~\cite{milletari2016v}              &   59.91      & 52.41      & 52.08      & \multicolumn{1}{c|}{54.80}    & 44.71         & 44.47         & 41.20      & 43.46      & 5.93       & 6.11       & 6.23       & \multicolumn{1}{c|}{6.09}     & 342.41     & 330.81        & 331.38     & 334.86 \\ 
                        & (1.02)       & (0.96)       & (0.80)       & \multicolumn{1}{c|}{(0.92)}     & (1.12)       & (2.99)       & (1.13)       & (1.69)     & (1.30)       & (1.20)       & (1.17)       & \multicolumn{1}{c|}{(1.22)}     & (2.27)       & (1.21)         & (1.08)       & (1.31)\\
                          \hdashline
                        SwinUNETR-V2~\cite{he2023swinunetr}     &   65.19      & 64.75      & 63.05      & \multicolumn{1}{c|}{64.33}    & 38.21         & 41.11         & 38.87      & 39.40      & 36.35      & 36.81      & 37.77      & \multicolumn{1}{c|}{36.98}    & 185.94     & 179.10        & 172.54     & 179.19 \\ 
                         & (2.03) &	 (3.25) &	 (2.31) &	 \multicolumn{1}{c|}{(2.53)} &	 (14.31) &	 (3.53) &	 (7.06) &	  (8.30) &	 (0.58) &	 (3.51) &	 (1.57) &	 \multicolumn{1}{c|}{(1.89)} &	 (4.13) &	 (1.01) &	 (1.45) &	 (2.20) \\ 
                          \hdashline
                        SwinUNETR~\cite{hatamizadeh2021swin}    &   67.02      & 67.57      & 64.93      & \multicolumn{1}{c|}{66.50}    & 34.10         & 37.41         & 35.06      & 35.52      & 36.86      & 38.29      & 35.61      & \multicolumn{1}{c|}{36.92}    & 177.88     & 168.84        & 177.53     & 174.75 \\ 
                         & (1.90) &	 (0.46) &	 (1.74) &	 \multicolumn{1}{c|}{(1.37)} &	 (4.62) &	 (6.76) &	 (5.96) &	 (5.78) &	 (0.79) &	 (0.66) &	 (0.08) &	 \multicolumn{1}{c|}{(0.51)} &	 (4.75) &	 (4.10) &	 (6.07) &	 (4.97) \\ 
                              \hdashline
                        UNETR~\cite{hatamizadeh2022unetr}       &   63.03      & 54.47      & 52.31      & \multicolumn{1}{c|}{56.60}    & 51.53         & 55.36         & 59.30      & 55.40      & 25.22      & 19.03      & 18.95      & \multicolumn{1}{c|}{21.07}    & 228.20     & 241.52        & 247.55     & 239.09 \\ 
                         & (1.20)       & (1.03)       & (0.99)       & \multicolumn{1}{c|}{(0.53)}     & (11.17)      & (12.55)      & (17.66)      & (13.56)    & (1.29)       & (3.00)       & (3.33)       & \multicolumn{1}{c|}{(1.75)}     & (8.55)       & (14.40)        & (16.61)      & (10.07)\\ 
                           \hdashline
                        nnFormer~\cite{zhou2023nnformer}        &   67.74      & 69.74      & 66.14      & \multicolumn{1}{c|}{67.87}    & 39.65         & 42.64         & 39.86      & 40.72      & 44.54      & 47.10      & 45.16      & \multicolumn{1}{c|}{45.60}    & 145.53     & 144.49        & 143.96     & 144.66 \\ 
             & (2.60) &	 (0.63) &	 (1.84) &	 \multicolumn{1}{c|}{(1.69)} &	 (5.66) &	 (0.42) &	 (2.19) &	 (2.76) &	 (0.74) &	 (0.34) &	 (0.08) &	 \multicolumn{1}{c|}{(0.38)} &	 (15.61) &	 (6.28) &	 (1.44) &	 (2.63)\\ 
                        \hdashline
                        SegResNet~\cite{myronenko20193d}        &   67.90      & 61.41      & 58.71      & \multicolumn{1}{c|}{62.67}    & 39.85         & 46.13         & 44.79      & 43.59      & 21.61      & 24.86      & 24.79      & \multicolumn{1}{c|}{23.75}    & 256.72     & 235.47        & 228.37     & 240.19 \\ 
                          & (0.19) &	 (0.31) &	 (0.27) &	 \multicolumn{1}{c|}{(0.13)} &	 (0.73) &	 (2.72) &	 (2.88) &	 (1.62) &	 (3.26) &	 (0.17) &	 (0.81) &	 \multicolumn{1}{c|}{(0.87)} &	 (14.55) &	 (3.03) &	 (14.22) &	 (0.90)\\
                                  \hdashline
                                                               
         MLP-UNEXT~\cite{li2024mlp}              &   69.46       & 61.71       & 60.29       & \multicolumn{1}{c|}{63.82}    & 28.71         & 40.51         & 39.44      & 36.22      & 18.92      & 17.95      & 17.20      & \multicolumn{1}{c|}{18.02}    & 267.22     & 261.28        & 265.72     & 264.74 \\ 
         & (0.21) &	 (1.09) &	 (1.43) &	 \multicolumn{1}{c|}{(0.91)} &	 (2.71) &	 (0.49) &	 (0.84) &	 (0.46) &	 (2.43) &	 (0.34) &	 (0.75) &	 \multicolumn{1}{c|}{(1.17)} &	 (15.36) &	 (3.84) &	 (3.56) &	 (7.58) \\
         \hdashline
        SegResNetVAE~\cite{myronenko20193d} & 70.97 & 69.04 & 62.37 & \multicolumn{1}{c|}{67.46} & 34.00 & 40.27 & 40.12 & 38.13 & 45.02 & 42.98 & 40.42 & \multicolumn{1}{c|}{42.81} & 151.60 & 154.93 & 152.37 & 152.97 \\
                     & (0.23) & (0.57) & (1.31) & \multicolumn{1}{c|}{(0.70)} & (0.20) & (2.23) & (2.05) & (1.36) & (0.70) & (0.17) & (0.06) & \multicolumn{1}{c|}{(0.31)} & (0.23) & (2.25) & (3.05) & (1.85) \\
         
    \midrule
    \rowcolor[HTML]{C8FFFD}
    Ours         &   \r{75.47} & \r{74.89} & \r{68.90} & \multicolumn{1}{c|}{\r{73.08}} & 
\r{15.32} & \r{20.31} & \r{19.99} & \r{18.54} & 
\r{47.87} & 49.30 & \r{46.90} & \multicolumn{1}{c|}{\r{48.02}} & 
\r{132.66} & \r{125.34} & \r{125.20} & \r{127.73} \\ 
    \rowcolor[HTML]{C8FFFD}
     (with full modality)                                      & \r{(1.71)} &	 \r{(0.32)} &	 \r{(0.65)} &	 \multicolumn{1}{c|}{\r{(0.46)}} &	 \r{(0.58)} &	 \r{(0.34)} &	 \r{(0.08)} &	 \r{(0.06)} &	 \r{(0.48)} &	 {(1.49)} &	 \r{(0.46)} &	 \multicolumn{1}{c|}{\r{(0.81)}} &	 \r{(8.24)} &	 \r{(0.29)} &	 \r{(2.77)} &	 \r{(3.57)}\\
                                           \hline
                                           
    \rowcolor[HTML]{C8FFFD}
    \textbf{Ours}                           &   \b{75.84}     & \b{75.25}    & \b{70.43}    & \multicolumn{1}{c|}{\b{73.84}}   & \b{13.96}        & \b{18.48}        & \b{18.27}    & \b{16.90}    & \b{53.32}    & \b{55.67}    & \b{51.94}    & \multicolumn{1}{c|}{\b{53.64}}   & \b{114.59}  & \b{111.35}      & \b{114.62}  & \b{113.52} \\
    \rowcolor[HTML]{C8FFFD}
     \textbf{(with t1c+t1+f)} & \b{(0.04)} &	 \b{(0.23)} &	 \b{(0.32)} &	 \multicolumn{1}{c|}{\b{(0.20)}} &	 \b{(3.82)} &	 \b{(0.41)} &	 \b{(0.46)} &	 \b{(0.67)} &	 \b{(1.15)} &	 \b{(0.78)} &	 \b{(0.84)} &	 \multicolumn{1}{c|}{\b{(0.92)}} &	 \b{(0.20)} &	 \b{(2.52)} &	 \b{(2.52)} &	 \b{(1.61)} \\
    \bottomrule
    \end{tabular}
    }
\label{tab:results}
\end{table*}
\subsubsection{Modality impact}
First, we analyzed the impact of each modality on the segmentation tasks using a 3D U-Net with a patch size of 128x128x128. \Cref{tab:results-modalities} summarizes the impact of each modality on segmentation performance. The differences in performance are due to the unique characteristics each modality captures.
T1 images offer high-resolution detail but low contrast between tissues, resulting in lower DSC scores. T1c, with a contrast agent, enhances visualization by increasing contrast between normal and pathological tissues, showing the highest performance among individual modalities. T2 images, sensitive to fluid content, are useful for detecting edema and show moderate performance. FLAIR imaging, which suppresses cerebrospinal fluid signals, enhances lesion visibility near the ventricles but may not differentiate tumor components as clearly as T1c.
Our results show that T1c excels in segmenting ET and NETC but is less effective for SNFH. T2 and FLAIR perform better in detecting SNFH. T1c is best for TC and ET, indicating fewer false positives and negatives, but it is not suitable alone for segmenting the entire tumor.
Combining two modalities shows that including T1c is crucial. T1c combined with T1 performs well in TC and ET segmentation, but overall performance improves when combined with T2 or FLAIR due to their complementary nature.
In three-modality combinations, T1c remains essential. To achieve the highest DSC scores, T1c should be combined with FLAIR. Adding either T1 or T2 to this combination yields good results, with T2 complementing FLAIR and T1 complementing T1c. The combination of T1c + FLAIR + T1 achieves the best average DSC in both legacy and lesion-wise scores.
Incorporating all four modalities leads to a decline in performance due to increased noise, confusion, redundancy, and information dilution. T2 and FLAIR overlap in highlighting fluid regions, and T1 and T1ce provide redundant structural and enhancing information. Increased noise arises from T2's sensitivity to fluid, and including multiple modalities increases the risk of artifacts and irrelevant signals. This results in suboptimal performance compared to three-modality combinations.

Since the combination of T1c, FLAIR, and T1 achieves the best average lesion-wise Dice score, we have selected these modalities for the two-stage segmentation of metastases.

\subsubsection{Two-stage model}
\Cref{tab:results} shows that our model improves segmentation performance on the BraTS Mets 2023 dataset compared to several state-of-the-art models that did not use any pre-training. We trained our two-stage segmentor from scratch with full modality and the three-modality combination that performed better in our modality impact study. Using the full modality, we achieved $\sim$1\% improvement in lesion-wise Dice score over the current state-of-the-art. However, by using the three modalities that had the most impact on lesion-wise metrics, we improved the overall results by $\sim$5\% in lesion-wise DSC and $\sim$0.8\% in legacy DSC.
It is noteworthy that CNN-based architectures like ResUNet and our model perform very well, while many transformer models struggle to achieve similar results. Our approach enhances segmentation accuracy and reliability, particularly in detecting small and challenging metastases. This improvement is evident in the better lesion-wise metrics. Moreover, our architecture is efficient in terms of FLOPS. We have only a 7 GFLOPS overhead from the detector, and due to our smaller input size compared to other architectures, we have lower FLOPS overall. For instance, while 3D-UNET requires about 478 GFLOPS, our approach reduces this to around 60 GFLOPS, an $\sim$85\% reduction. Additionally, since we do not segment any patch that the detector indicates does not contain metastases, we only use 7 GFLOPS, significantly reducing calculations. This is beneficial since most patches do not contain metastases, and we gain from this efficiency.
\begin{figure}[!t]
    \centering
    \resizebox{\textwidth}{!}{
        \begin{tabular}{@{} *{6}c @{}}
            \includegraphics[width=0.18\textwidth]{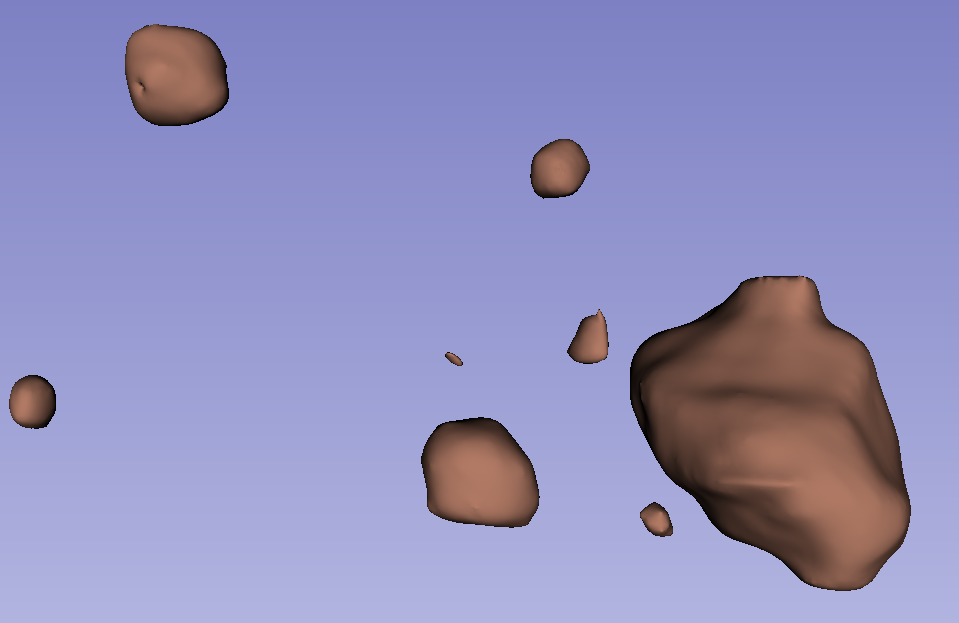} &
            \includegraphics[width=0.18\textwidth]{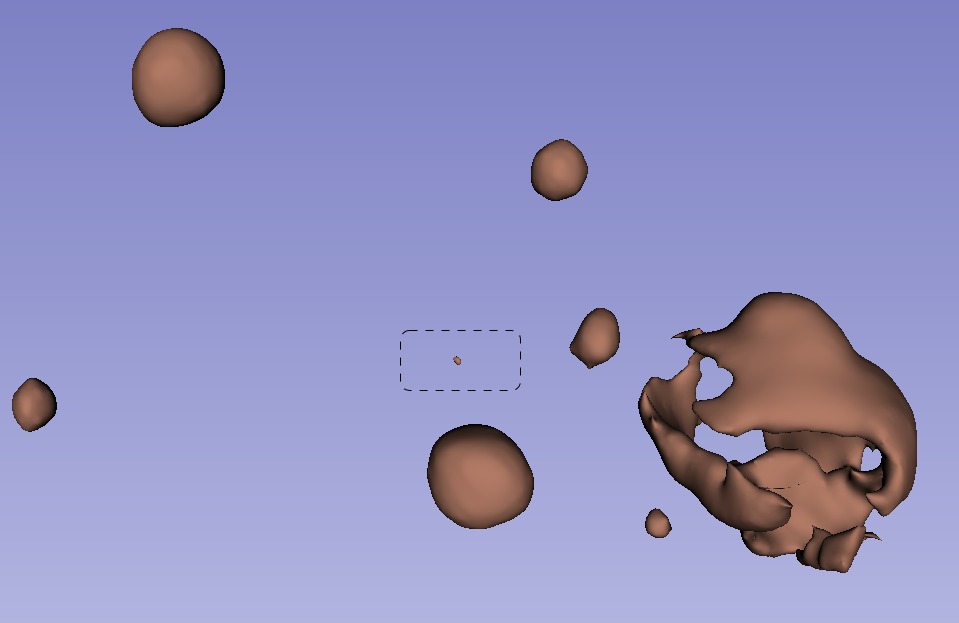} &
            \includegraphics[width=0.18\textwidth]{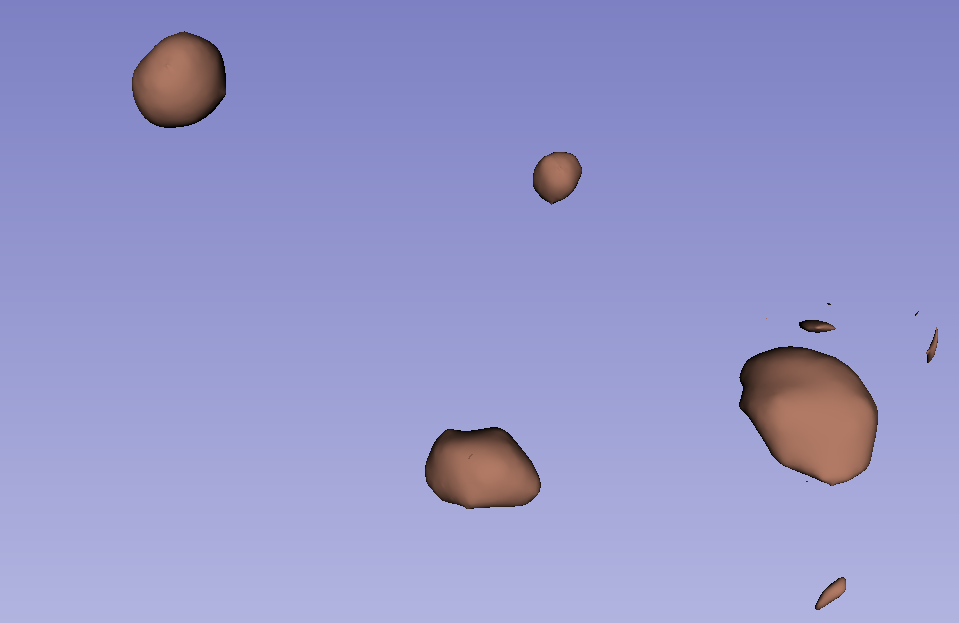} &
            \includegraphics[width=0.18\textwidth]{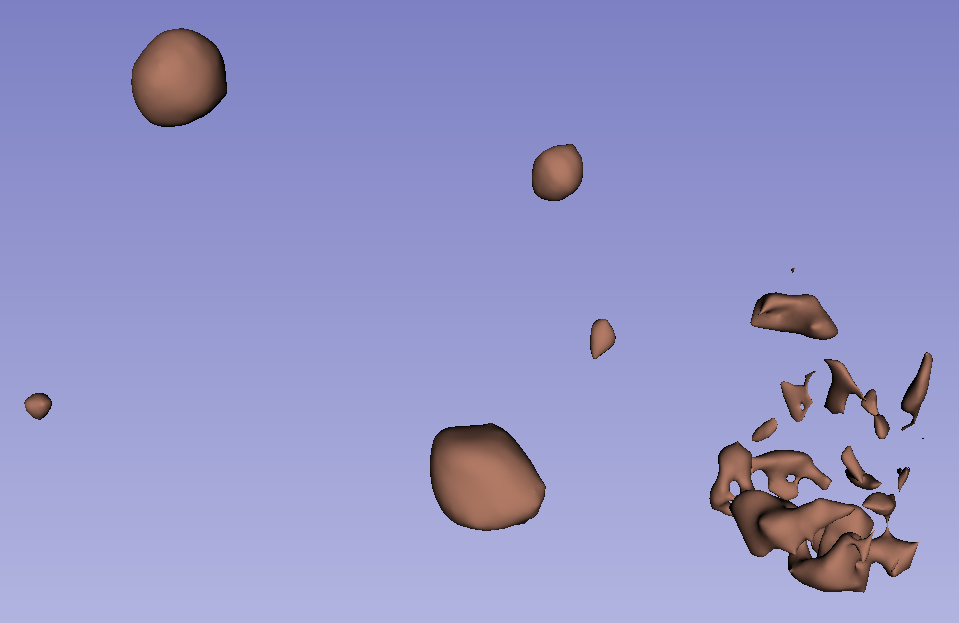} &
            \includegraphics[width=0.18\textwidth]{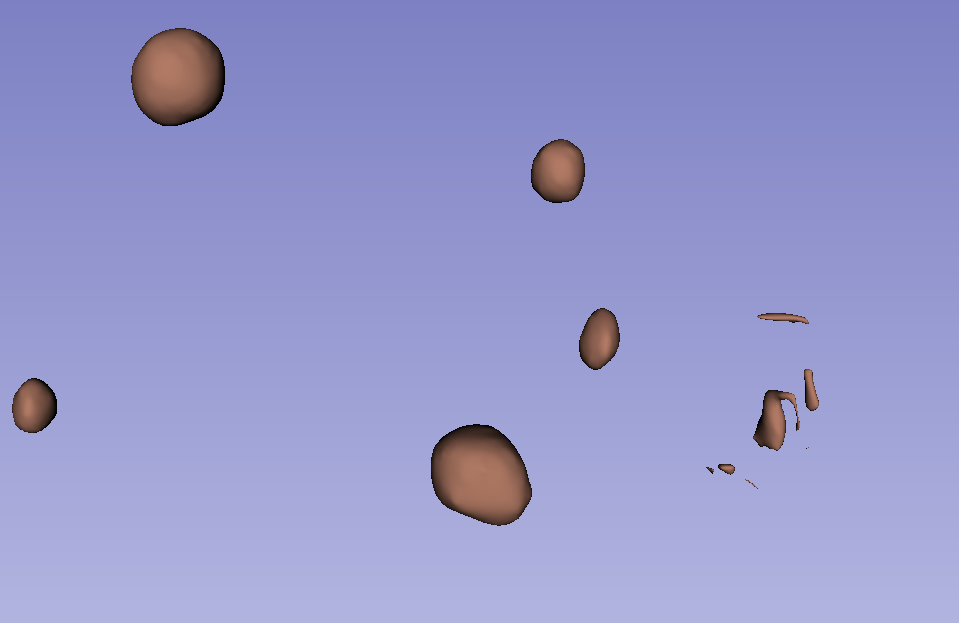} \\
            \includegraphics[width=0.18\textwidth]{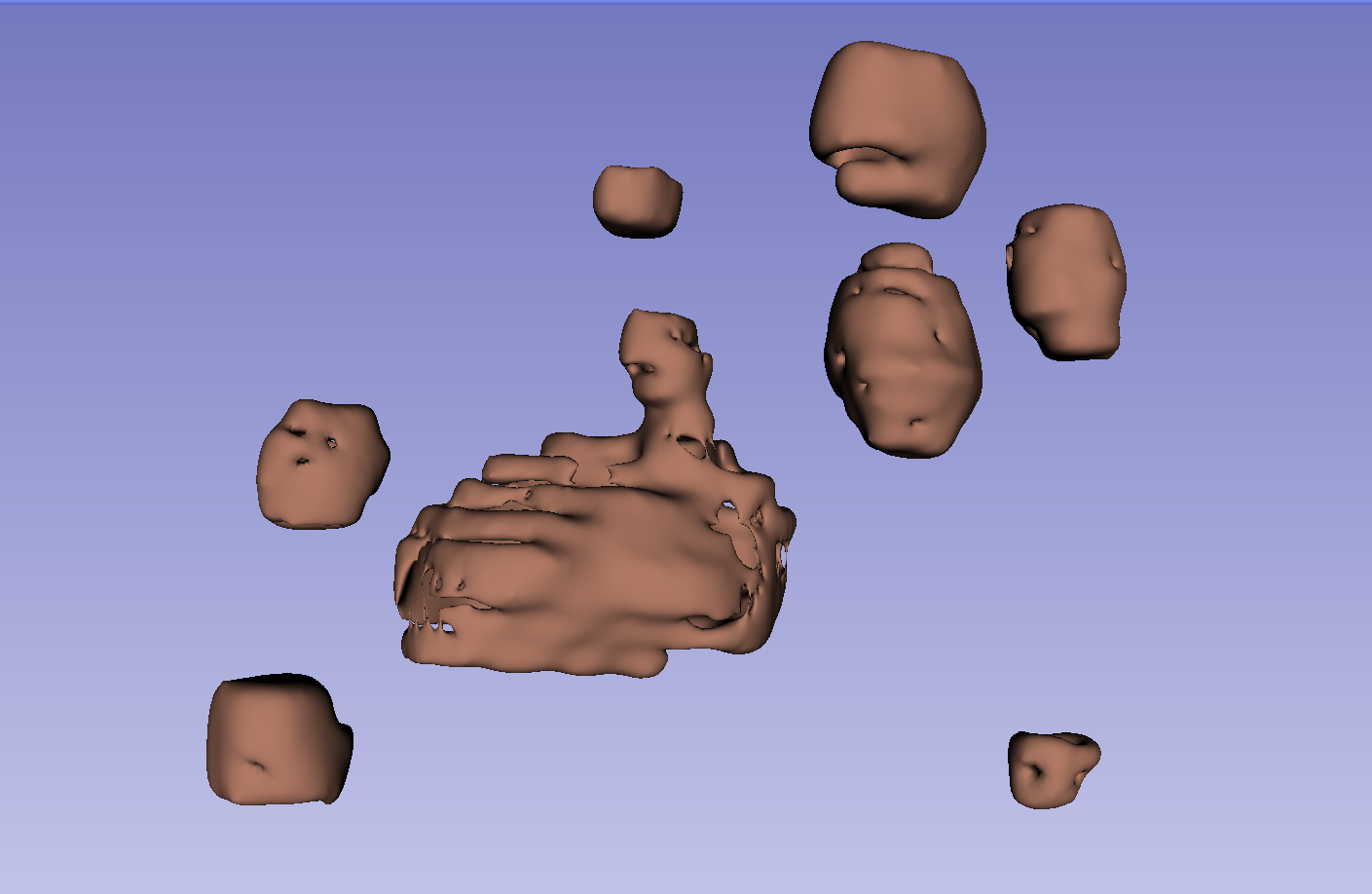} &
            \includegraphics[width=0.18\textwidth]{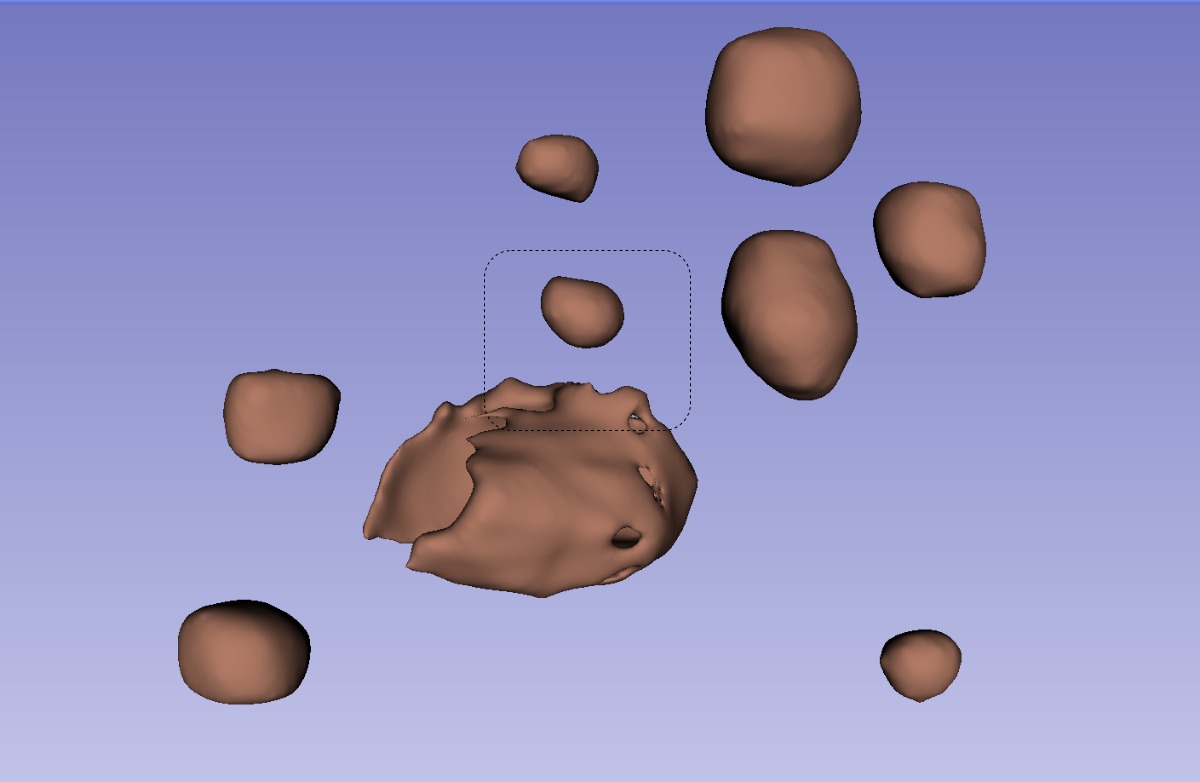} &
            \includegraphics[width=0.18\textwidth]{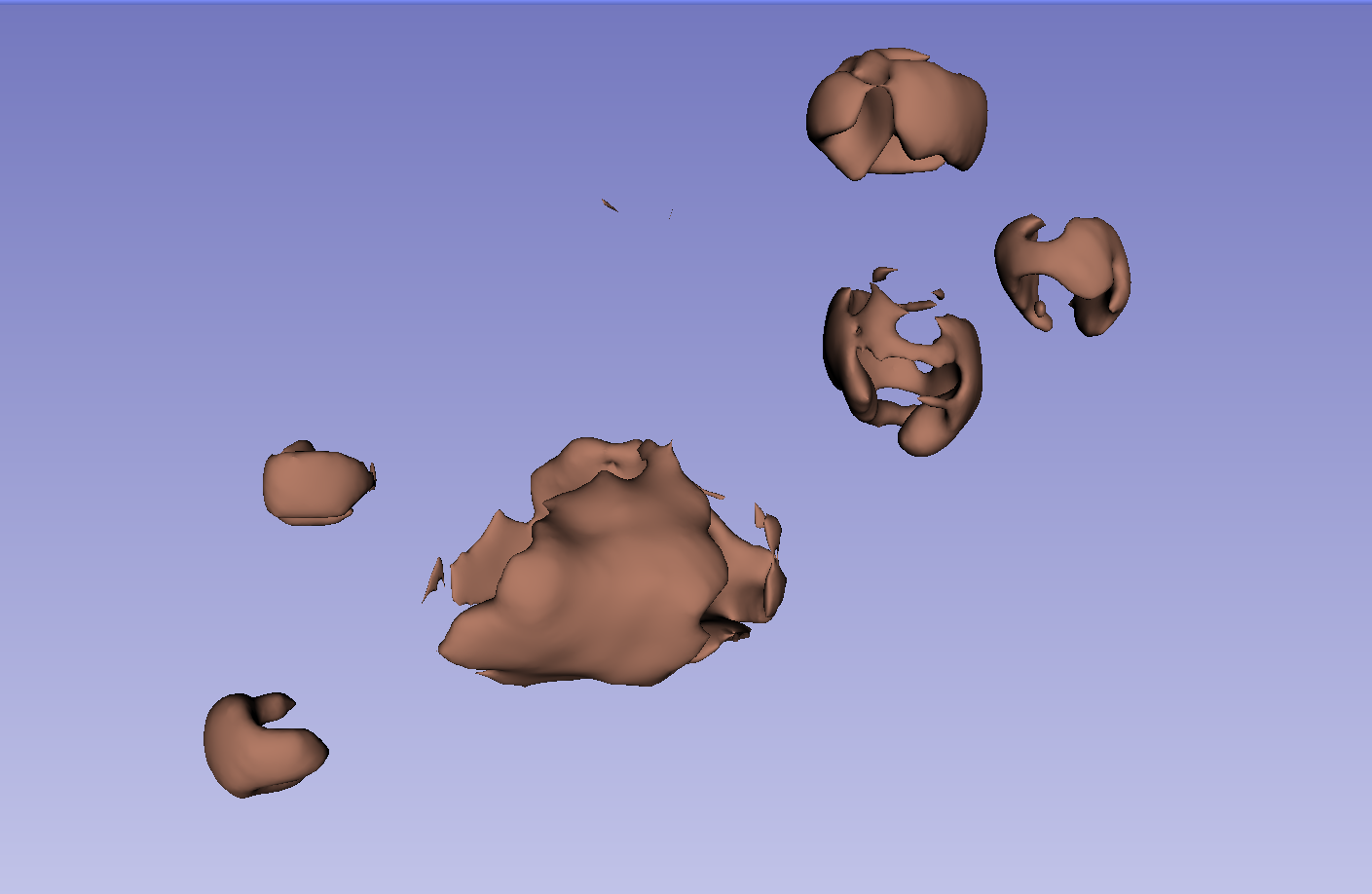} &
            \includegraphics[width=0.18\textwidth]{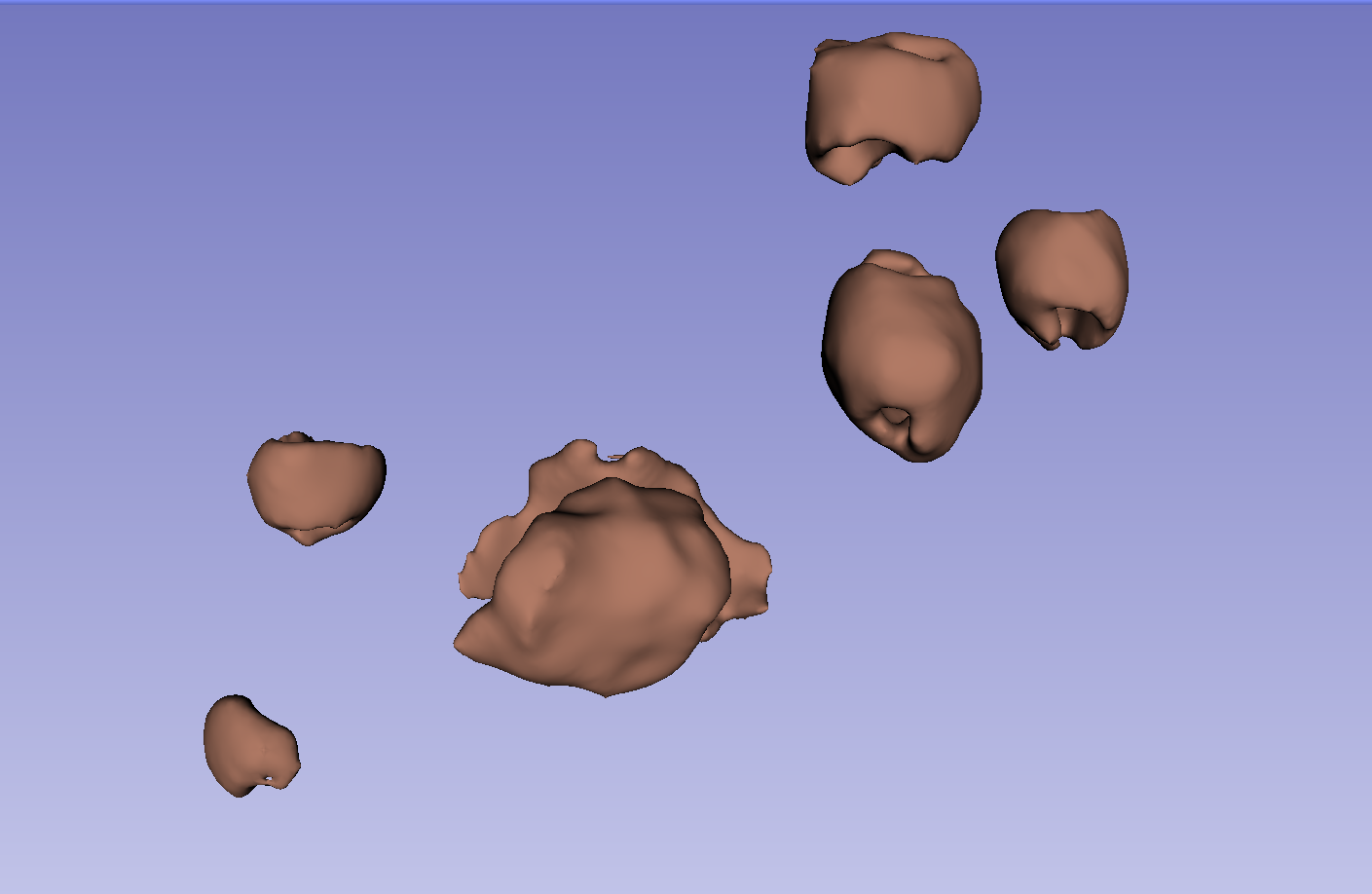} &
            \includegraphics[width=0.18\textwidth]{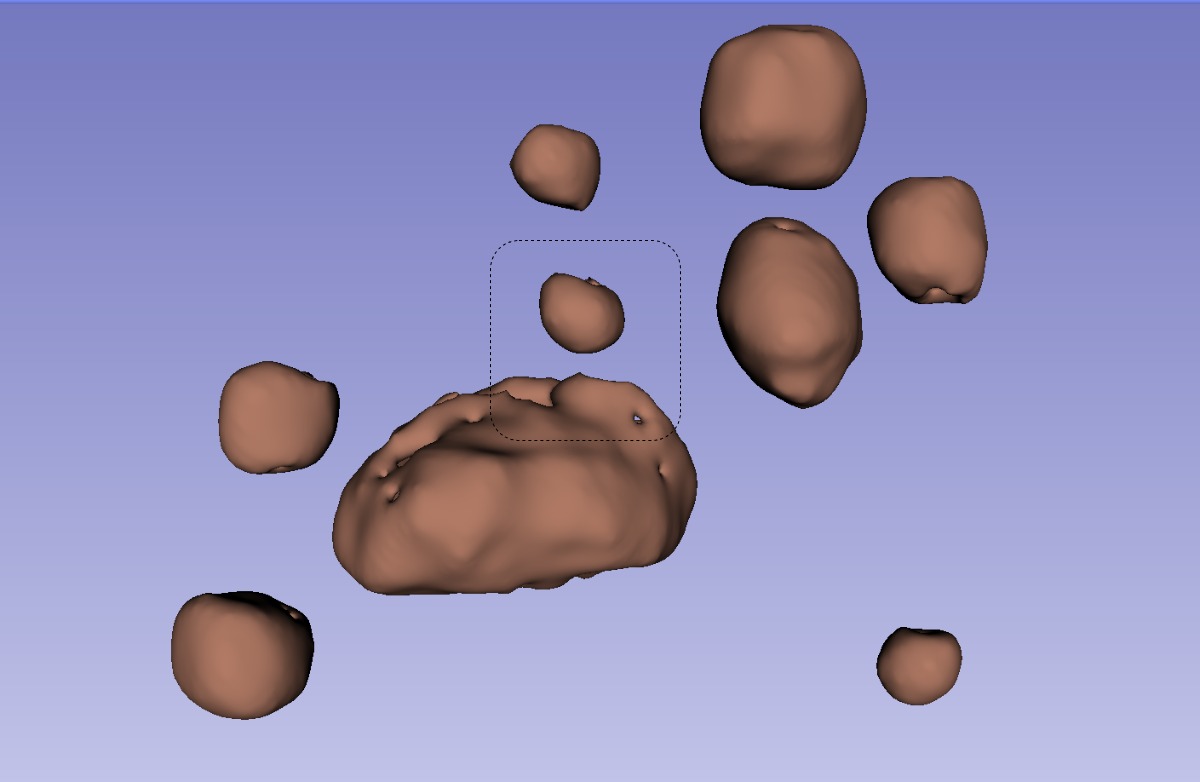} \\
            {\small Ground Truth} & {\small Ours (t1+t1c+f)} & {\small UNETR}  & {\small SwinUNETR-V2} & {\small nnFormer}
        \end{tabular}
    }
    \caption{Visual comparisons of different methods on the BraTS-Mets 2023 dataset (images were generated using 3D Slicer~\cite{Fedorov_3D_Slicer_as_2012}).} 
    
    \label{fig:visualcomparison}
\end{figure}

\subsection{Qualitative results}
\label{subsec:qualitative}
\Cref{fig:visualcomparison} compares the 3D segmentation results of metastases from various models against the ground truth. Our model accurately segments both large and small metastases. In the first row (patient), it captures the smallest lesion in the centre, unlike other models. In the second row (patient), it effectively detects and segments the available metastasis. These results highlight the superior performance of our model in accurately detecting and segmenting brain metastases compared to other models.


\section{Conclusion}
\label{sec:conclusion}
In this paper, we introduced a two-stage detection and segmentation model for brain metastasis. Our study on the impact of different imaging modalities revealed that combining T1c, T1, and FLAIR modalities yields the most accurate segmentation results. Notably, using all available modalities did not enhance performance. Our proposed model significantly improves lesion-wise Dice scores compared to other single-pass models, highlighting its potential for more precise metastasis diagnosis and treatment. This research underscores the importance of strategic modality selection and multi-stage processing in achieving superior segmentation performance.

\section*{Acknowledgemnet}
\label{sec:ackknowledgemnet}
\sloppy
The authors gratefully acknowledge the scientific support and HPC resources provided by the Erlangen National High-Performance Computing Center (NHR@FAU) of the Friedrich-Alexander-Universität Erlangen-Nürnberg (FAU) under the NHR project “DeepNeuro - Exploring novel deep learning approaches for the analysis of diffusion imaging data.” NHR funding is provided by federal and Bavarian state authorities. NHR@FAU hardware is partially funded by the German Research Foundation (DFG) – 440719683.


%
%

\bibliographystyle{splncs04}
\bibliography{28.bib}
%




\end{document}